\documentclass[12pt,preprint]{aastex}

\newcommand{\cmiii}{\mbox{cm$^{-3}$}}

\newcommand{\ha}{\mbox{\sc{H}{$\alpha$}}}
\newcommand{\hi}{\mbox{H\,{\sc i}}}
\newcommand{\hii}{\mbox{H\,{\sc ii}}}

\newcommand{\kms}{\mbox{km~s$^{-1}$}}

\newcommand{\radmii}{\mbox{rad~m$^{-2}$}}
\newcommand{\pccmvi}{\mbox{pc~cm$^{-6}$}}

\slugcomment{To appear in {\it The Astrophysical Journal}}

\shorttitle{The Fragmenting Superbubble Associated with W4}
\shortauthors{West, English, Normandeau, and Landecker}

\received{2005 June 23}

\begin{document}

\title{The Fragmenting Superbubble Associated with the \hii\ Region W4}

\author{Jennifer L. West}
\affil{Department of Physics and Astronomy, University of Manitoba  Winnipeg, Manitoba, Canada, R3T 2N2}
\email{Jennifer\_West@UManitoba.ca}

\author{Jayanne English}
\affil{Department of Physics and Astronomy, University of Manitoba  Winnipeg, Manitoba, Canada, R3T 2N2}
\email{english@physics.umanitoba.ca}

\author{Magdalen Normandeau}
\affil{Department of Physics, Amherst College, 
       Amherst, MA, 01002, United States of America\footnote{Present address,
Department of Physics, University of New Brunswick, PO Box 4400, 8 Bailey Drive, Fredericton New Brunswick, Canada, E3B 5A3}}
\email{mnormandeau@amherst.edu}

\and

\author{T.\ L.\ Landecker}
\affil{Dominion Radio Astrophysical Observatory,
       Herzberg Institute of Astrophysics,
       National Research Council,  
	P.O.\ Box 248, Penticton, B.C., V2A 6J9, Canada}
\email{Tom.Landecker@nrc.gc.ca}

\begin{abstract}

New observations at high latitudes above the \hii\ region W4 show that
the structure formerly identified as a chimney candidate, an opening
to the Galactic halo, is instead a superbubble in the process of
fragmenting and possibly evolving into a chimney. Data at high
Galactic latitudes ($b > 5\arcdeg$) above the W3/W4 star forming
region at 1420 and 408 MHz Stokes I (total power) and 1420 MHz Stokes
Q and U (linear polarization) reveal an egg-shaped structure with
morphological correlations between our data and the \ha\ data of
Dennison, Topasna, \& Simonetti.  Polarized intensity images show
depolarization extending from W4 up the walls of the superbubble,
providing strong evidence that the radio continuum is generated by
thermal emission coincident with the \ha\ emission regions. We
conclude that the parts of the \hii\ region hitherto known as W4 and the
newly revealed thermal emission are all ionized by the open cluster
OCl 352. Assuming a distance of 2.35 kpc, the ovoid structure is ~164
pc wide and extends ~246 pc above the mid-plane of the Galaxy.  The
shell's emission decreases in total-intensity and polarized intensity
in various locations, appearing to have a break at its top and another
on one side. Using a geometric analysis of the depolarization in the
shell's walls, we estimate that a magnetic field line-of-sight
component of 3 to 5 $\mu G$ exists in the shell. We explore the
connection between W4 and the Galactic halo, considering whether
sufficient radiation can escape from the fragmenting superbubble to
ionize the kpc-scale \ha\ loop discovered by Reynolds, Sterling \&
Haffner.  
\end{abstract}


\section{Introduction}
\label{sec:intro}

Theory suggests that supernovae and/or winds from massive stars
existing in clusters confined to the disk of the Galaxy produce
bubbles in the interstellar medium which expand, becoming
superbubbles, and burst out of the Galactic disk producing collimated
outflows into the Galactic halo (Norman \& Ikeuchi 1989; Tenorio-Tagle
\& Bodenheimer 1988). The superbubbles, which create cavities in the
atomic hydrogen gas in the disk of the Galaxy, contain a plasma of hot
gas, high energy photons, and cosmic rays (CRs).  When a bubble's wall
bursts at high latitudes the structure is termed a ``chimney'' since
the cavity and its walls act as a conduit for the outflow of this
plasma towards the halo of the Galaxy (see e.g.\ de Avillez \& Berry
2001). In addition, the magnetic field that is oriented parallel to
the disk of the Galaxy is ``frozen'' into the plasma and is thus
``pulled'' upwards with the shell material as the bubble
expands. Consequently, it is predicted that these conduits should have
magnetic field lines that run tangential to the shell (Norman \&
Ikeuchi 1989; Basu, Johnstone \& Martin 1999, hereafter BJM99).

An unusual, cone shaped void observed in neutral hydrogen (Normandeau,
Taylor \& Dewdney 1996, hereafter NTD96), which has the star cluster OCl
352 at its base, is one of the few observationally identified Galactic chimney
candidates and has come to be known as the ``W4 Chimney''. It was
proposed that the region has been evacuated of \hi\ by the winds of 9
O-type stars that are part of the open cluster OCl 352. In this case
the stellar wind scenario is favoured over the supernovae scenario
since the cluster seems too young to have had any supernovae occur and
the energy output of the winds appears to be sufficient to account for
the void (NTD96). Estimates of the age of OCl 352 range from 1.3--2.5
Myr (Dennison, Topasna \& Simonetti 1997, hereafter DTS97, and
references therein) to 3.7--4.3 Myr (NTD96). We adopt a distance to
the region of 2.35 kpc.

The ``Chimney'' region is located above OCl 352 and is a distinct
component from the ionized ``W4 loop'', which is located below OCl 352
(Normandeau, Taylor \& Dewdney 1997). We consider only the structure
that is above the star cluster and above the ridge of ionized gas that
defines the upper latitude structure of the W4 HII region. The HI
images (see NTD96 and Figures~\ref{fig:HI-43.40_general} and
\ref{fig:HI-43.40_faint}) make it clear that the regions above and
below the cluster, i.e. above and below the upper ionized ridge of W4,
are different in nature: the upper region has been mostly cleared of
HI, hence the possibility that it is a chimney or a superbubble. The
lower region has not been cleared, at least not at the same
velocities.

The interpretation of Terebey et al. (2003) and BJM99 that the
superbubble structure includes the W4 loop is in contrast to this
picture. While it is likely that both the W4 loop and the structure
above W4 are powered by the same stars, they do not form one dynamical
entity. The use of the term ``chimney'' provides a metaphor only for the
part of the structure above W4 since a chimney not only has walls but
it also has a fire at its base. In this case the fire is the OCl 352
cluster. To avoid unwieldy nomenclature, rather than referring to the
structure as the W4 superbubble/chimney, we dub it G134.4+3.85. The
new name, however, should not be understood as implying that
G134.4+3.85 has a different origin from W4. Indeed we will argue that
both objects are thermal emission regions excited by the open cluster
OCl 352.

In the radio r\'egime, at Galactic latitudes of less than +4\arcdeg,
the structure is reminiscent of Norman \& Ikeuchi's (1989)
conventional chimney model, i.e.\ a broken \hi\ shell allowing CRs and
UV photons to escape to higher latitudes. The contrasting analysis of
DTS97, using narrow-band \ha\ data, is that W4 is an ``apparently
closed superbubble''. However the DTS97 data suffer from substantial
vignetting at the field edges which could lead the eye to see a bubble
shape where none exists. This makes corroboration from other datasets
desirable. Additionally, the DTS97 data do not include velocity
information, which means that the observed \ha\ emission could be from
unrelated sources along the line of sight.

BJM99 considered both the dynamics (Kompaneets modelling) and the
ionization of the superbubble based on the \ha\ data from DTS97. Of
particular interest here is the conclusion of BJM99 that an open geometry in
\hi\ images and a closed one in data showing ionized gas are not
mutually exclusive. The superbubble's shell could be sufficiently thin
at high latitudes that, while it closes the shell and prevents
streaming of gas towards higher latitudes, it does not hold back the
ionizing radiation which then obliterates the \hi\ at higher
latitudes. This scenario is in fact likely according to these
authors.

In this paper we present new observations of radio continuum and HI emission, extending coverage from $b = 5\arcdeg$, the limits of earlier data, to $b = 8.5\arcdeg$. We analyze the \hi\ spectral line emission as well as the 1420 MHz and
408 MHz radio continuum emission from this region for evidence
supporting either the chimney or the superbubble picture. For the 1420
MHz continuum data, full polarization information is available. The
data acquisition and processing are described in
\S\ref{sec:data}. Image mosaics and false colour morphological
analysis are presented in \S\ref{sec:mosaics}. The \hi\ emission is
addressed in \S\ref{sec:HI}. The polarization data are discussed in
\S\ref{sec:polar}. The inter-relationship between emission components is discussed in \S\ref{sec:inter}  and the magnetic field of the region is explored,
using our polarization data, in \S\ref{sec:Best}.

Our analysis creates a picture of a {\em fragmented} superbubble,
G134.4+3.85, above W4 that may be in the process of evolving into a
chimney possibly due to magnetic effects. While we use two separate names we believe that both
``objects'' are the outcome of the ionizing flux from OCl 352, and are
therefore in some sense just one object. In \S~\ref{sec:discuss} we
discuss this picture, superbubble formation scenarios, and the
superbubble's connection to the Galactic halo. The paper is summarized
in \S~\ref{sec:conclu}.

\section{Data}
\label{sec:data}

\subsection{Acquisition}
\label{sec:acquisition}

Landecker et al.\ (2000) provide a complete description of the
Dominion Radio Astrophysical Observatory (DRAO) Synthesis
Telescope. The Telescope is an east-west array of seven antennae with
diameters $\sim$9 m. Of these seven, four are fixed and three are
movable allowing sampling of baselines from 12.86 m to 604.29 m in
steps of 4.29 m.  At $\lambda = 21$ cm, this corresponds to coverage
on all spatial scales from $\sim$1 arcmin to 56 arcmin.  The data
collected consist of several components: spectral line emission from
atomic hydrogen at $\lambda = 21.1$ cm, continuum emission at 1420 MHz
with full polarization information, and continuum emission 408 MHz.

The DRAO Synthesis Telescope recently completed Phase I of the
Canadian Galactic Plane Survey (CGPS; Taylor et al. 2003). The CGPS is
a relatively high resolution ($\sim$1\arcmin\ at 1420 MHz;
$\sim$3.7\arcmin\ at 408 MHz) dataset covering 73.1\arcdeg\ in
longitude, from $\ell = 74.2\arcdeg$ to $\ell = 147.3\arcdeg$ along
the Galactic plane of the Milky Way.  The latitude extent covers
$-3.6\arcdeg < b < +5.6\arcdeg$.  Three higher latitude fields, on an
extension of the CGPS grid, were acquired for the present study.

At 1420 MHz the field of view of the Synthesis Telescope has a full
width at half maximum (FWHM) of 107.2\arcmin. At 408 MHz, the FWHM of
the field of view is 332.1\arcmin. The survey region has been observed
on a regular hexagonal grid with 112\arcmin\ between field
centers. The 1420 MHz and 408 MHz mosaics used in the
present study consist of 11 CGPS fields plus the additional 3 fields
observed at latitudes greater than +5.6\arcdeg\ having field centers
135.47\arcdeg, +5.85\arcdeg; 134.53\arcdeg, +7.46\arcdeg; and
133.60\arcdeg, +5.85\arcdeg. The polarization mosaics contain the 14
fields described above plus an additional 4 CGPS fields, which
reveal more of the ``ambient'' Galactic medium surrounding the W4
region. \hi\ mosaics consist of the pilot project data (Normandeau,
Taylor \& Dewdney 1997) plus 3 additional CGPS fields.

Since the minimum spacing of the DRAO Synthesis Telescope is 12.86 m,
data from single-antenna telescopes are added to regain missing large
scale structure. At 1420 MHz, single-antenna continuum,
total-intensity data are obtained from the 100 m Effelsberg Telescope
(Reich et al.\ 1990, 1997) for latitudes up to +4\arcdeg, and from the
Stockert 25-meter northern sky survey (Reich 1982, Reich \& Reich
1986) for higher latitudes. The Effelsberg Telescope has a resolution
of 9\arcmin, whereas the Stockert Telescope has a resolution of
35\arcmin. The Effelsberg data were absolutely calibrated using the
Stockert survey data, which ensures consistency amongst the
contributions from the short-spacing data. Short-spacing information is
only added to the 1420 MHz total intensity continuum maps (Stokes I)
and not to the Stokes Q or Stokes U polarization maps. At 408 MHz,
short-spacing data are obtained from the all sky map of Haslam et al.\
(1982) with a resolution of 51\arcmin. For the spectral line data,
short-spacing information was collected using the DRAO 26-m Telescope.

The radio continuum emission at 1420 MHz is observed in four discrete
7.5 MHz bands centered at $\pm 6.25$ MHz and $\pm 13.75$ MHz from the
\hi\ spectral line frequency of 1420.406 MHz. This leaves a 5 MHz gap
at the central frequency, which is sufficient to avoid contamination
from the \hi\ spectral line. The central 1 MHz of the gap is used for
the spectral line observations. The four continuum bands are combined
when producing the images used in this paper (Stokes I, Stokes Q and
Stokes U).

Taylor et al.\ (2003) indicate that there is an uncertainty in the
absolute scale of the 408 MHz data of up to $\sim$15\% due to
uncertain calibration data. The internal precision of the data is
unaffected by this error.

\subsection{Reduction and Processing}
\label{sec:reduc}

The processing of all CGPS fields and the addition of short-spacing
data in all fields were carried out by members of the CGPS data
processing team using custom software developed at DRAO (Willis 1999;
Taylor et al. 2003). Processing of the additional, high-latitude
fields and the mosaicking of fields were carried out by the authors
using the DRAO software, with the assistance of the CGPS data
processing team. The higher latitude data were processed, calibrated,
and supplemented with short-spacing data such that they matched the
CGPS survey data.

\section{Results}
\label{sec:results}

\subsection{Mosaics}
\label{sec:mosaics}

Mosaics showing the \hi\ spectral line emission at velocities of
interest are shown in Figures~\ref{fig:HI-43.40_general}, 
\ref{fig:HI-43.40_faint}, \ref{fig:HI-40}, and \ref{fig:HI-48}.  The mosaics of continuum emission at 408
MHz and 1420 MHz, Stokes I, Q and U are presented in
Figures~\ref{fig:408} through \ref{fig:QU}. The grating rings that remain visible in the HI and polarization data are due to the high intensity sources W3 and 3C~58. The 1420 MHz data were
convolved to match the lower resolution ($\sim$3.7\arcmin) of the 408
MHz data.

The polarization data were convolved to a resolution of 5\arcmin\ to
improve the signal to noise ratio and the Stokes Q and U images were
combined to produce a polarized intensity image ($PI = \sqrt{Q^2 +
U^2}$) and a polarization angle map ($\psi =
\onehalf\arctan\left(\frac{U}{Q}\right)$) (See
Figure~\ref{fig:PIPang}). Measurements of Stokes I intensities and PI
for 56 regions of interest are tabulated in West~(2003).

\subsection{\hi}

\label{sec:HI}

\subsubsection{The atomic hydrogen structure of G134.4+3.85}

\label{sec:HI_descrip}

Figure~\ref{fig:HI-43.40_general} shows the \hi\ data at $v_{\rm LSR}
= -43.40$ \kms.  The eastern \hi\ wall of the superbubble is clearly
visible up to +5.5\arcdeg, at which point it curves slightly inward
and disappears. The western wall is only well defined up to
+3.4\arcdeg.  In the new data set presented here, there is no
evidence of a cap at high latitudes in the \hi. At lower latitudes,
the cavity spans 5 channels and LSR velocities --38.46 to --45.05
\kms. We consider the region defined by $\rm 133.55 < l < 135.5$, $\rm
6\arcdeg < b < 7\arcdeg$, the latitudes associated with the \ha\ shell
described by DTS97. We assume that the cap would have a velocity width of 5 channels (8 \kms), the velocity width of the cavity. Averaging the signal over 5 channels, we could have detected a cap of column density $4 \times \ 10^{19}$~$\rm{cm^{-2}}$. Using 5 channels reduces $\sigma$ from the value of the rms noise in a single channel, which is 3.3~K or $1.6 \times \ 10^{19}$~$\rm{cm^{-2}}$.  Thus if such a cap had been detected its column density would be at the 6$\sigma$ level. The observed column density in the top of the ionized bubble is $1 \times \ 10^{19}$~$\rm{cm^{-2}}$.  By
comparison, the column density of electrons in the DTS97 \ha\ shell is
$\rm \sim 1 \ \times \ 10^{19}$~cm$^{-2}$.  For this latter estimate
we converted emission measure into column density using an average of
DTS97's measurements of 4 points in the upper shell, T = $10^4$~K,
their emission measure conversion of 1R = 2.8~pc~cm$^{-6}$, and
their shell thickness estimate of 10 to 20 pc.  If there was neutral
material associated with this \ha\ shell at a density comparable to
that of the ionized material, we would not have been able to detect
it.

There is a small filament of \hi\ pointing upward, away from the plane
(Figure~\ref{fig:HI-43.40_faint}). This low level feature is present in four channels
of the mosaic, from --41.76 \kms\ to --46.70 \kms.  A line through this linear feature passes through OCl~352. We will argue in \S\ref{sec:chimney} that this is material within the superbubble
which remains neutral, protected from the ionizing radiation by
shadowing.  This faint arc is well below the latitude of
the cap claimed by DTS97 and therefore below the upper boundary of the
best fit Kompaneets model found by BJM99.  Furthermore, the upper tip
of the vertical filament is above the ionized cap suggested by the DTS97 data.

It is noticeable that the OCl 352 cluster is at the base of the \hi\
cone, as was described by NTD96, whereas the Kompaneets model by
BJM99, which was fit primarily to the H$\alpha$ image, shows a bubble
extending to lower latitudes, to the base of the ionized gas loop
which forms the lower half of W4.

This conical shape is not unique - something similar is seen at the
base of the Aquila supershell (Maciejewski et al.\ 1996). The W4 cone extends
approximately 200 pc upward from the OCl 352 cluster, based on the
well-defined upper longitude wall. This is comparable to the cone at
the base of the Aquila supershell which extends roughly 175 pc for an
assumed distance of 3.3 kpc.

\subsubsection{Scale height in the vicinity of G134.4+3.85}

\label{sec:scale_height}

The most surprising result from the modeling by BJM99 is the
implication of a very small scale height ($H$), namely 25 pc. This
value was obtained by matching the aspect ratio of the superbubble as
seen in H$\alpha$ using the Kompaneets model and assuming a distance
of 2.35 kpc. 

Using the data from the 26-m Telescope, which extends further up in
latitude, almost to +10\arcdeg, and fitting an exponential to the
decay in column density, a scale height of 140 $\pm$ 40 pc is found for
the \hi\ in the vicinity of the G134.4+3.85 (c.f.~Figure~\ref{fig:Nvsz}). A variety of different velocity intervals were used to test the robustness of this result, as well as different data
sets (26-m data, CGPS pilot project data, and CGPS data), and all
yield a scale height of approximately 140 pc, which is significantly larger 
than the 25 pc required by the model of BJM99. This value compares well
with the values of 135 pc quoted by Lockman, Hobbs \& Shull (1986).

Apart from the contradiction with the prediction by BJM99, this result
is not particularly surprising: scale heights for the neutral medium
are in the 100-200 pc range. Clearly, however, although the fit of the
Kompaneets model to the shape of the W4 superbubble is very good, it
does not address all of the physical processes occurring in this
vicinity. In particular, magnetic fields may have contributed to the
collimation (Komljenovic et al. 1999, de Avillez \& Breitschwerdt 2005 (see \S\ref{sec:time})).

It should be noted that the scale height evaluated here used data from
all longitudes covered by the available 26-m data, from 124.4\arcdeg\
to 144.4\arcdeg\ with varying coverage in latitude. However, if only
longitudes greater than those corresponding to G134.4+3.85 are used,
the profile is greatly different for latitudes below +6\arcdeg,
particularly between +1.5\arcdeg\ and +4.0\arcdeg\ where the high
longitude data show a plateau instead of a decrease. The difficulty
lies in deciding which profile is representative of the medium into
which the superbubble expanded. The effect of including G134.4+3.85
itself in the longitude range is minor compared to the effect of what
appears to be a partial shell at lower longitudes which may be
associated with 3C~58. These difficulties generate the uncertainty of
$\pm$ 40 pc associated with the 140 pc scale height used in this
paper.

\subsection{Polarization}
\label{sec:polar}

Gray et al.\ (1999) presented polarization data from the low latitudes
($b < +4\arcdeg$) of this region. The high electron density combined
with the magnetic field in the W3/W4/W5 region act as a ``Faraday
screen'', causing the polarized, background Galactic synchrotron
radiation passing through the plasma to be rotated by the process of
Faraday rotation. These effects are observationally recognized as
depolarized regions that follow the contours of the \hii\
regions. Depolarization occurs when the polarization angle varies by a
significant amount between vectors contributing to a single
measurement.  The polarization angle can differ between adjacent lines
of sight sampled by a single telescope beam (beamwidth depolarization)
or vary along the path length (depth depolarization). A third
depolarization mechanism, bandwidth depolarization, becomes important
if the rotation measure (RM) is very high at the observing wavelength,
and so the angle may change by a large amount across the observing
band.  Bandwidth depolarization is not significant in this study as typical RMs in the
region are on the order of $|100|$ \radmii\ and to produce significant
depolarization over the 35 MHz observing band of these observations, a
RM of $\sim$1400 \radmii\ would be required (Brown 2002).

In Fig.~\ref{fig:QU} and Fig.~\ref{fig:PIPang} we see that the
depolarization extends without interruption from W4 up the ``walls''
of G134.4+3.85. This is seen in Stokes Q and U and in polarized
intensity. The continuity of depolarization from W4 to the walls of
the superbubble provides strong evidence that the superbubble is
connected to the HII region; this is not a chance superposition. A
temperature spectral-index map (West 2003) also suggests that W4 and
G134.4+3.85 are a single coherent structure.

\subsubsection{The wishbone}
\label{sec:wishbone}

An unusual wishbone-shaped feature is visible in both Stokes Q and
Stokes U as well as in polarized intensity and polarization angle
maps. This feature has no counterpart in Stokes I,
total-intensity. The feature appears above G134.4+3.85 with
approximate center coordinates (134.9\arcdeg, +7.1667\arcdeg) (see
Fig.~\ref{fig:QU} and Fig.~\ref{fig:PIPang}). The morphology of the
object suggests a shell-like structure. A local steepening of the
spectral index is observed south of the wishbone (West 2003).
 
\subsubsection{Comma-shaped Polarization Feature}
\label{sec:comma}

A bright, comma-shaped knot of strong polarized intensity with
approximate center coordinates of (134.5\arcdeg, +2.58\arcdeg) is
particularly prominent in polarized intensity (Fig.~\ref{fig:PIPang})
but invisible in total-intensity. It is $\sim$0.5\arcdeg\ wide and
$\sim$0.2\arcdeg\ high and has a peak polarized intensity of 0.46
K. The object appears near the base of the prominent V-shaped
filaments which are visible in both \hi\ and 60 $\mu$m IRAS dust
emission. However, there is no clear morphological evidence of an
association.

The SIMBAD database was searched for catalogue objects which might be
coincident with the comma-shaped knot on the plane of the sky. A
prominent object in this region is an O8 star, BD+62 424 with
coordinates 134.53\arcdeg, +2.46\arcdeg. There have been two,
distinctly different heliocentric radial velocity measurements for
this star: Abt, Levy, \& Gandet (1972) measured a value of --42.5 \kms
(--40.2 \kms LSR) while Fehrenbach et al.\ (1996) measured a value of
--9 \kms (--6.7 \kms LSR). The measurement by Abt et al.\ is
consistent with the chimney velocities as observed in the \hi\
spectral line. However, the radial velocity measurement of Fehrenbach
et al. (1996) yields a kinematic distance of $\sim$800 pc (Brand \& Blitz
1993) implying that this star is a foreground object.  The high
polarized intensity of this knot suggests that it may be closer than the
W4 region as it suffers less depolarization than the surrounding region.

\section{Interrelationship Between Emission Components}
\label{sec:inter}

To aid in comparison of the various wavelengths, a colour image
combining the datasets is shown in Figure~\ref{fig:colour}. This
image illustrates that there is correlation of radio continuum and
\ha\ emission, but that the IRAS emission seems uncorrelated. The
\hii\ regions W3, W4 and W5, the supernova remnants (SNRs) HB3 and
3C~58, as well as the spiral galaxy Maffei 2 are readily seen. The \ha\
superbubble, identified by DTS97, is much more apparent in these
combined colour images than in the individual images. In
Fig.\ref{fig:labelled} we outline an egg-shaped bubble and suggest its
three-dimensional structure. A feature that seems to be a continuous
branch of emission coming off the main shape of G134.4+3.85 has been
labeled the fork.

The new data presented in this paper show strong evidence of radio
continuum emission above OCl 352, associated with the W4 loop below
the star cluster, and extending up to a latitude of $b \approx
+6.7\arcdeg$.  The emission seen at 1420 MHz and 408 MHz generally
confirms the \ha\ dataset of DTS97, and reveals G134.4+3.85 to have a
striking, continuous egg-shaped structure. In addition, the
polarization data reveal a depolarized region consistent with the
shell-like structure and indicate that thermally emitting material is
depolarizing the background synchrotron emission (see
\S\ref{sec:polar} and \S\ref{sec:Best}). On the east wall we see
clearly that the ionized material (whose position is judged from the
depolarization feature in Figures~\ref{fig:QU} and \ref{fig:PIPang})
lies inside the neutral material (position judged from
Figure~\ref{fig:HI-43.40_general}). The relative temperature spectral
index is flatter within the region outlined by the depolarized shell
and the value of the index indicates that this emission is thermal
(West 2003). 

Although the data suggest that G134.4+3.85 has a top (we have already
noted in \S\ref{sec:HI} that our observations do not have sufficient
sensitivity to show atomic gas connected with the ionized material at
the top of the shell), the shape does not have a uniform intensity
around the perimeter. There are regions of lower intensity that
suggest breaks in the continuous structure.  Two main breaks appear in
the structure, one near the top and one on the lower longitude side
near $b = +4\arcdeg$.  These are seen at both radio frequencies and in
\ha\ with closely coincident positions. The center coordinates and
approximate widths for the two breaks are listed in
Table~\ref{tbl:breaks}.

At 1420 MHz and in \ha, there is emission, nicknamed the fork, that
branches off the main structure (Figure~\ref{fig:contours}). A faint
arc of emission extends up from the high longitude side of the fork
and appears to arc westward back to G134.4+3.85. This arc of emission
is coincident with the top portion of the polarization wishbone. The
lower portion of the fork exhibits depolarization similar to that of
the lower portion of the wall, suggesting that the fork is associated
with G134.4+3.85. However it is unclear whether the
continuing arc of emission is associated with G134.4+3.85.  A hand
tracing of the regions of higher emission including the fork and the
apparent breaks is presented in Figure~\ref{fig:contours}.

There is a distinctive region of very low continuum emission intensity
inside the walls of G134.4+3.85 above $b \approx +3.5\arcdeg$. It
extends up to +4.8$\arcdeg$ and spans $134\arcdeg < \ell <
135.5\arcdeg$. Figure~\ref{fig:Ivsb_Ivsl} shows plots of the 1420 MHz
data at fixed longitude ($\ell = 134.68\arcdeg$) and fixed latitude
($b = +4.011\arcdeg$). In both plots this region of very low emission
($< 4.8$ K) is evident; in Figure~\ref{fig:Ivsb_Ivsl}a the area runs
from $+3.5\arcdeg < b < +4.8\arcdeg$ and in
Figure~\ref{fig:Ivsb_Ivsl}b the area extends from $134\arcdeg < \ell <
135.5\arcdeg$.  The region has lower emission than even the regions
exterior to G134.4+3.85. This low emission region is most obvious at
1420 MHz but is also apparent in both 408 MHz and \ha. There
appears to be 60 $\mu$m emission in this region from dust that one
could reasonably expect to find in the foreground along the line of
sight. The low emission implies either that the region is evacuated of
plasma that would emit at these wavelengths or that, in the case of the
\ha\ data, UV photons contained within are being absorbed by dust. In
the former case, there would be lower plasma column density along the
lines of sight passing through this region producing a lower observed
intensity.

In this low emission region, there are two horizontally oriented \hi\
filaments positioned at $b \approx +3.75\arcdeg$ which were first
identified by NTD96. These are either foreground or background
features, although, judging by their radial velocity, quite likely to
be in the vicinity of G134.4+3.85. They are unliekly to be within it,
given the evidence that ionization extends to much higher latitudes.

Additionally, a ridge of \ha\ and 1420 MHz emission appears at $b
\approx +3\arcdeg$ to +3.5\arcdeg. The ridge at 1420 MHz is visible in
Figure~\ref{fig:StokesI}. Coincident with the 1420 MHz ridge is a
flattening of the temperature spectral index and below this ridge but
above DTS97 point ``O'' (West 2003) the spectral index steepens,
indicating a shift between these two features (West 2003).  This
supports the picture of a lower region containing electrons and
ionizing radiation (at $b < +3.75\arcdeg$) with an evacuated region
above the ridge (at $b > +3.75\arcdeg$).

To summarize, although the structure appears open in the \hi\ images,
data showing the ionized gas, along with polarization and temperature
spectral index maps, delineate an egg-shaped bubble with a shell that
appears to be broken in at least two locations.  The
temperature spectral index map and \ha\ data suggest that radiation
may possibly be constrained to the lower part of the bubble by a ridge
that is evident in \ha\ and 1420 MHz data.

\section{Magnetic Field Estimate}
\label{sec:Best}

Rotation measures of extragalactic sources seen through the bubble
were measured in an attempt to determine the magnetic field in the
shell \citep{west2003} but the uncertainties are large and the number
of sources is small, allowing $B_\parallel$ to take any value between
1 and 24$\mu$G.  Since rotation measure analysis does not constrain
the magnetic field well, we use the following approach based on
depolarization in the walls of the shell.  We noted above that the
walls of G134.4+3.85 are seen in our data as regions of low
polarization relative to the surroundings of the superbubble and that
they join smoothly to the depolarization region coinciding with W4
\citep{polar}. From the observed degree of depolarization in the walls
we can derive an estimate of the magnetic field in the shell of the
superbubble.  We use the electron density derived from the \ha\
measurements of DTS97, and base our calculation on an assumed geometry
of the walls.

The electron density is derived from the \ha\ flux measurements of
DTS97 measured at four locations along the upper arc ($b > +5\arcdeg$)
of the shell (DTS97's points labeled H, I, J, and K; see
Figure~\ref{fig:contours}). We estimate the shell thickness, $s$, from
the projected intensity to be between 10 and 20 pc. Assuming a
temperature of 10$^4$ K, and using 1 R (Rayleigh) as equivalent to an
emission measure of EM = 2.8 \pccmvi, and DTS97's result that the
emission measure of the shell $\sim5 n_e^2 s$, then $n_e$ is
calculated to be 0.44 $\pm$ 0.19 \cmiii\ for an assumed shell
thickness of 10 pc, or 0.31 $\pm$ 0.13 \cmiii\ for an assumed shell
thickness of 20 pc. The range of possible electron density is
approximately ${0.2}<{n_e}<{0.6}$ \cmiii.

For the following geometrical argument, we assume that the superbubble
is cylindrical, with its axis perpendicular to the Galactic plane (and
therefore perpendicular to the line of sight). Its cross-section is
circular, with outer diameter 80 pc and wall thickness between 10 and
20 pc. The line of sight passes through the cylinder of ionized
material in which the electron density is assumed to take the
values derived above, and the polarized synchrotron background
emission will suffer Faraday rotation on passage through this material
if a magnetic field is present and has a line-of-sight component. The
path length through the wall of the cylinder for adjacent lines of
sight will differ, and so the amount of Faraday rotation will
differ. If the rotation is sufficiently different for adjacent paths
within the telescope beam, there will be significant beam
depolarization. The data presented in Figures~\ref{fig:QU} and ~\ref{fig:PIPang}, where the
angular resolution is 5\arcmin\,, indicate that the depolarization
amounts to $\sim$50\%, implying that the Faraday rotation on lines
through the shell separated by 5\arcmin\ differ by about
60$^{\circ}$. At the distance of G134.4+3.85 (2.35 kpc) 5\arcmin\
corresponds to a physical distance of 3.5~pc. A simple geometrical
calculation shows that two paths through the center of the wall of the
cylinder, separated by a distance 3.5~pc transverse to the line of
sight, will differ in length through the wall by ${\Delta}{l}={13}$~pc
if the wall thickness is 10~pc and by ${\Delta}{l}={19}$~pc if the
wall thickness is 20~pc.

The Faraday rotation is

\begin{equation}
{\Delta\theta}={0.81{\lambda^2}{n_e}{B_{\parallel}}{{\Delta}{l}}}
\label{eq:RM}
\end{equation}
where $\lambda$ is the wavelength, $n_e$ is the electron density,
$B_{{\vert}{\vert}}$ is the line-of-sight component of the magnetic
field, and ${\Delta}{l}$ is the path length.  Inserting the values
${\Delta\theta}={60^{\circ}}$, ${\lambda}={0.21\thinspace{\rm{m}}}$,
with values for ${n_e}$ and ${\Delta}{l}$ derived above, we find
${3.4{\mu}{\rm{G}}}<{B_{\parallel}}<{9.1{\mu}{\rm{G}}}$. The field
derived for the ``average'' parameters (${n_e}={0.38}$ \cmiii\ and
${\Delta}{l}={15}$~pc) is ${B_{\parallel}}={5.0{\mu}{\rm{G}}}$.

The range of possible values of electron density in the walls of the
superbubble, ${0.2}<{n_e}<{0.6}$ \cmiii\ is 10 to 30 times the ambient
electron density in the vicinity of W4, $\sim$0.02 \cmiii\ 
(Taylor \& Cordes 1993). This implies that the material in the walls of
G134.4+3.85 has been compressed by this factor. $B_\parallel$ at this
longitude is expected to be $\sim{1}{\mu}{\rm{G}}$, and $B_{total}$ is
about ${4}{\mu}{\rm{G}}$ (Beck 2001). If $B$ varies as ${n}^{0.5}$,
then we expect ${B_{\parallel}}$ in the walls of the bubble to be between
3.1 and ${5.5}{\mu}{\rm{G}}$, compatible with the above field
estimate.

However, we need to consider the likelihood that magnetic field and
electron density will be uniform, as is assumed in our field
calculation.  In the Q, U, and PI images (Figures~\ref{fig:QU} and ~\ref{fig:PIPang}) the
surroundings of G134.4+3.85 present a mottled appearance that is
strongly suggestive of a non-uniform magneto-ionic medium. If we think
of the medium as divided into ``cells'' in which the field and
electron density are more or less constant, then the size of the cells
appears to be about 20\arcmin\ in the surroundings. With a compression
factor of 10 to 30, the linear dimension of the cells will shrink by
the square root of the compression factor to about 4\arcmin. Beam
depolarization, which would arise from averaging across many cells, is
then unlikely to be significant in the 5\arcmin\ data, and the
geometrical effect we have proposed seems to be an adequate
description of the processes leading to the observed depolarization.

Nevertheless, there is still a problem. At 1\arcmin, the full
resolution of the observations, we would expect to see much less
depolarization, both when the geometrical effect is dominant and when
the bubble walls contain many cells, and we should not see the bubble
walls as significant depolarization features. However, at 1\arcmin\
resolution the eastern wall of the bubble is still detectable as a
depolarization feature (although the western wall is not). There must
be some cells in the eastern wall as small as 1\arcmin, but we cannot
place a precise value on the extent of this small-scale structure
because the signal-to-noise ratio at 1\arcmin\ is very low. The level
of polarized intensity in the eastern wall relative to that of the
surroundings is very roughly 1.6, although the uncertainty in this
ratio is so high that we cannot claim that it is higher or lower than
the value of 2.2 $\pm$ 0.3 measured from the data at 5\arcmin\
resolution.

In conclusion we note that BJM99 propose the presence of a swept-up
magnetic field, tangential to the shell of the W4 chimney/superbubble,
that potentially has a stabilizing effect on the structure (by
suppressing Rayleigh-Taylor instabilities that tend to break up the
shell). Kolmjenovic et al. (1999) use magnetohydrodynamic modeling to
argue that a {\it{vertical}} magnetic field of a few microGauss is
required to explain a bubble structure that is so highly collimated,
and to provide the required stabilizing effect. Our depolarization
observations indicate that a line-of-sight component of 3 to 5
${\mu}{\rm{G}}$ exists in the shell. However, this measurement does
not exclude the possibility of a stronger vertical component, giving a
total field of 4 to 7 ${\mu}{\rm{G}}$ or larger.

\section{Discussion}
\label{sec:discuss}

\subsection{ Superbubble or chimney?}
\label{sec:chimney}

At our adopted distance of 2.35 kpc, the projected ellipse of
G134.4+3.85, which measures $\sim$4\arcdeg\ wide by $\sim$6\arcdeg\
high, has physical width $\sim$164 pc and height $\sim$246 pc.  It has
risen to almost twice the scale height of the \hi\ , $\sim$140 pc,
determined in \S\ref{sec:scale_height}, and at that level the \hi\
density is only 13\% of its mid-plane value
(Figure~\ref{fig:Nvsz}). Nevertheless, the bubble has not reached the
halo, considered to begin at a height of $\sim$500 pc
\citep{Lockmanetal86}.  Two-dimensional numerical simulations
(e.g. ~\nocite{MMN89}MacLow, McCray \& Norman 1989) and analytic
theoretical explorations (e.g.~\nocite{FerTol00}Ferrara \& Tolstoy
2000) indicate that at a distance away from the plane of about 3 scale
heights a bubble will accelerate and break out into the halo, creating
a chimney.  Not only has G134.4+3.85 not reached this height, it may
not have the high levels of input wind (or supernova) energy to do so.
Most treatments of a single superbubble evolving in a uniform
environment indicate that substantially more driving energy is
required for breakout. For example, MacLow, McCray, \& Norman (1989)
show that a bubble driven by a mechanical luminosity of $1.67 \times
10^{38}$ erg~s$^{-1}$, more than five times the output of OCl 352,
will break out into the halo. Tomisaka (1998) describes the evolution
of a bubble driven by $3 \times 10^{37}$ erg~s$^{-1}$, equal to the
mechanical luminosity of OCl~352, and show that it is unlikely to
break out, and, furthermore, is stabilized by the ambient magnetic
field.

McClure-Griffiths et al.\ (2003) describe the Galactic chimney
GSH~277+00+36, perhaps a counter-example to G134.4+3.85. This is
clearly a chimney, very empty inside, with very complex HI structures
in the walls, interpreted as hydrodynamic instabilities. In this case
the gas flow within the chimney has had a major impact on the neutral
gas forming its walls. We see no such features in
G134.4+3.85. Furthermore, GSH~277+00+36 reaches to a height of 1~kpc
above the Galactic plane, at least four times as high as
G134.4+3.85. G134.4+3.85 is either driven by a smaller mechanical
energy than GSH~277+00+36, or is at a much earlier stage of evolution.

Thus the data we have presented seem at first glance to support a
picture in which G134.4+3.85 is a superbubble, not a chimney. The
structure of ionized gas tapers at the top, suggesting convergence or
closure.  The 1420 MHz and \ha\ data clearly show this converging top,
and, to a lesser degree, so do the 408 MHz data.  There is no cap of
neutral material at the upper levels in our \hi\ data. Nor is there any detectable
neutral shell surrounding the ionized one; rather G134.4+3.85 appears as a cavity in the surrounding medium.  The \hi\ images do,
however, show the diverging conical shape exhibited by chimney models
(Norman \& Ikeuchi 1989).  These observations can be understood in
terms of the picture presented by BJM99. In the upper reaches of the
bubble the ambient density is dropping rapidly and the ionization
structure is unbounded. All neutral gas has been ionized. In this
picture, any neutral gas found within the conical void must be
protected from UV photons by shadowing.  The prime example is the
pair of filaments forming a {\bf{V}} at $b \approx 2.7\arcdeg$. A
second possible instance, found in our observations, is the vertical
\hi\ filament extending from $\sim$(134.15\arcdeg, +6.083\arcdeg) to
$\sim$(134.175\arcdeg, +7.25\arcdeg) and with a width
$\sim$0.25\arcdeg\ (see Figures~\ref{fig:HI-43.40_faint} and
\ref{fig:labelled}), which appears in several velocity channels from
--38.46 \kms\ to --45.05 \kms, velocities which are consistent with
G134.4+3.85 velocities. These features are linear, or made up of
straight-line elements, and are oriented in line with the position of
OCl~352.

Although the ionized shell appears closed, our observations have also revealed
two gaps; details are given in Table~\ref{tbl:breaks}. These regions
of low emission may well be breaks in the shell. One break is at the
top of the bubble, at $b = 6.3\arcdeg$, and has an apparent width of
20~pc. The ambient density is lowest at the top of the shell, and this
is exactly where one might expect the shell to be disrupted. The other
break is at lower latitude, and appears to be larger, nearly 50~pc
wide. In this break we see neither neutral or ionized material.
We interpret these breaks as evidence that the superbubble is
fragmenting, and turn to an existing theoretical treatment of
superbubble breakout and disruption to see what might be learned from
it. 

We base our discussion on the recent work of de Avillez \&
Breitschwerdt (2005), who have made high-resolution, three-dimensional
simulations of the magnetized ISM. Their calculations trace in fine
detail the evolution of such a medium driven by supernova explosions
at the Galactic rate. While they start with a simple, uniform medium,
their simulations follow the evolution into the more realistic
situation of a medium that becomes highly processed as the result of
stellar activity. The morphology of G134.4+3.85 is quite similar to
some superbubbles shown in their Figure 2. A bubble has developed,
centered at $x\approx0.2$kpc, $z\approx0.25$kpc, in which the higher
density skin, rather than forming a closed shell, is open at the top
and can be described as filamentary or fragmented. Given its density,
this skin can be identified as cool \hi\ gas if the region is in
pressure equilibrium.  Interior to this skin is a more continuous
lower density lining which could be identified with emission from
warmer ionized gas.  This bubble has protrusions of even hotter gas at
both the top and side which would appear as breaks in ionized gas
observations, remarkably similar to the breaks seen in our continuum
data for G134.4+3.85. The dimensions of the bubble are approximately
200 pc laterally and 600 pc vertically, somewhat larger than
G134.4+3.85

Because of this height, the model bubble can be considered to have
reached the Galactic halo at ${z}\approx{500}$~pc.  It has been shown
with 2-D simulations (e.g.~\nocite{MMN89} MacLow, McCray \& Norman
(1989)), that, at this altitude of a few scale heights,
Rayleigh-Taylor instabilities (in which hot gas pushes into cooler,
denser gas) will occur and cause fragmentation of the shell.  However,
instabilities, Rayleigh-Taylor or other types, can occur at earlier
stages, according to the 3-D simulations. Indeed the dense skins of
smaller bubbles in the de Avillez and Breitschwerdt simulation are
also filamentary, indicating that this is so.

Although a magnetic field in the shell can stabilize it against
Rayleigh-Taylor instabilities, de Avillez and Breitschwerdt mention
that MHD instabilities can occur which will mix the magnetized
shell with the ambient medium and cause the escape of hot gas.  The
field in the skin of their model bubble is a few $\mu$G or less, and
Rayleigh-Taylor instabilities would not be suppressed by this
field. The ambient field around G134.4+3.85 is only a few $\mu$G,
but, due to compression, the ionized lining of G134.4+3.85 could
have a magnetic field strength as high as 7 $\mu$G (\S\ref{sec:Best}).
Thus Rayleigh-Taylor instabilities may be reduced and MHD
instabilities, instead, may be the cause of both breaks apparent in
G134.4+3.85. The simulations need to be examined in detail in
order to assess whether this is the case; de Avillez has begun such an
investigation in relation to G134.4+3.85 (private communication).

In summary, G134.4+3.85 has now risen into the lower density levels of
the Galactic disk, but not yet into the halo. Its ionized
shell is apparently closed indicating that it is a
superbubble.  At high latitudes, the \hi\ distribution retains the
conical chimney structure, but that is because the ionized shell is
transparent to ionizing radiation.  Gaps seen in the shell may
well be the first signs of fragmentation of the superbubble and
indicate that hot gas is escaping. If so, G134.4+3.85 is on
its way to becoming a chimney. The process is likely to accelerate as
supernova explosions among the stars of OCl~352 inject energy into the
bubble.

\subsection{Timescales and possible formation scenarios}
\label{sec:time}

W4 and its immediate neighbour, W3 (IC~1795), together comprise a
complex, highly processed sector of the Galaxy.  Many authors have
suggested that multiple epochs of star formation have occurred in the
region (Lada et al. 1978; Thronson et al. 1980; Carpenter, Heyer \&
Snell 2000; Reynolds, Sterling \& Haffner 2001).  In particular, Oey
et al. (2005) offer evidence that IC~1795 was actually triggered by W4
and that IC~1795 has recently triggered the formation of three
ultracompact \hii\ regions around W3.  Given that this process is
still underway, a question arises: is G134.4+3.85 simply the product
of the winds from OCl~352, or has it been shaped by earlier activity?
The axis of G134.4+3.85 passes through the position of OCl~352, and
there can be little doubt that this star cluster is now driving the
evolution of the bubble. However, if winds from earlier massive stars
swept out the environment, G134.4+3.85 could have extended to its
present height relatively quickly.  Oey et al. argue further that the
very low scale height ($\sim$25~pc) deduced by BJM99 is the result of
earlier clearing of the environment by a long-gone generation of
stars. However, the exact alignment of OCl~352 with the axis of
G134.4+3.85 suggests that the superbubble is not simply filling an old
cavity, which would most probably have no such alignment.

NTD96 calculate that the combined winds of the 9 O-type stars in the
cluster could blow out the region up to a height of 110 pc in 5.7 Myr
which roughly agrees with cluster age estimates of 3.7 to 4.3 Myr. In
light of \ha\ evidence that the structure extends up to $\sim$230 pc,
DTS97 calculate that 6.4 to 9.6 Myr would be required to allow
sufficient expansion. They cite cluster ages of 1.3 to 2.5 Myr, smaller
than those obtained by NTD96, and thus claim that OCl 352 cannot be
solely responsible for the formation of the \ha\ superbubble.
In passing we note that both the calculations of NTD96 and DTS97 ignored
magnetic fields. Ferri{\`{e}}re et al. (1991) have shown that a bubble
expands more slowly in a magnetized medium: an even longer time
would be needed to reach the present size of G134.4+3.85.

Reynolds et al.\ (2001) observe an \ha\ loop extending 1300~pc above
the midplane of the Galaxy and centered on coordinates (139\arcdeg,
+15\arcdeg), which they believe is being ionized by OCl~352.
They also suggest that mutiple star formation epochs have been involved
in the formation of this loop, and suggest a timescale of 10 to 20
Myr for the process. If OCl~352 is maintaining the ionization of
material 1300~pc above the plane, then the region must be completely
clear of neutral material. This is confirmed by our observations, at
least to a height of about 300 pc.

We support the suggestion that OCl~352 is the dominant source of
ionizing radiation for G134.4+3.85 but that it is not solely
responsible for the creation and evolution of the
structure. Further, we postulate that there is little matter
remaining interior to the upper portion of G134.4+3.85 and that any
gas that may have existed at an earlier stage of the life of the
superbubble has escaped through the apparent breaks in the structure
of G134.4+3.85.

We have presented evidence that there is significant magnetic field in
the walls of G134.4+3.85 (\S\ref{sec:Best}). Ferri{\`{e}}re et
al. (1991), Tomisaka (1998), and de Avillez \& Breitschwerdt (2005)
all consider bubbles in the magnetized ISM; all conclude that bubbles
elongate in a direction parallel to the field. Tomisaka (1998)
concludes that the field will inhibit expansion perpendicular to the
field, and can also contain the expansion of the bubble. The expansion
can still accelerate vertically, even in the presence of a field
aligned with the Galactic plane, but only if it is driven by enough
energy to reach several scale heights.

What then do we make of the {\it{vertical}} elongation of G134.4+3.85?
Either it indicates that the scale height is extremely low, as
suggested by BJM99, or it implies a field component orthogonal to the
Galactic plane (as suggested by Komljenovic et al. 1999). Since we do
not believe that the scale height is as low as the 25~pc derived by
BJM99, we must consider the possibility of a vertical component to the
field.  Such a component may be the relic of an earlier bubble and we
are compelled by the very shape of G134.4+3.85 to consider that it may
have formed in a pre-processed medium, not a pristine one. We can
envisage a situation where a new bubble has filled an old one, and the
elongation may have become stronger with each successive wave of star
formation. This fits well into the picture of triggered star formation
for this complex (Oey et al. 2005) where the new generation of stars is
triggered within the walls of the superbubble created by the preceding
generation.

There is also useful insight to be gained from analytical
approximations available in the literature. Both Ferri{\`{e}}re et
al. (1991) and de Avillez and Breitschwerdt (2005) derive analytic
equations applicable to a magnetohydrodynamic medium.

De Avillez \& Breitschwerdt (2005) calculate the age and the
radius (in the equatorial plane) of the bubble when its expansion
stalls.  Their expression for the age at stall is
\begin{equation}
{t_{max}}={{7.2 \times
10^{2}}\thinspace{L_{37}^{1/2}}\thinspace{n_{0}^{3/4}}
\thinspace{B_{-6}^{-5/2}}\thinspace{\rm{Myr}}}
\end{equation}
where ${L_{37}}$ is the mechanical luminosity of the stellar wind in
units of $10^{37}$ erg{\thinspace}s$^{-1}$, $n_0$ is the density in
cm$^{-3}$, and $B_{-6}$ is the ambient magnetic field in units of
$\mu$G. The derivation of this expression is based on the assumption
of a uniform ISM and a regular magnetic field. The bubble stalls when
the force from thermal pressure is balanced by magnetic force as the
bubble distorts field lines. Inspection of equation 2 shows that the
stall time is very strongly dependent on the magnetic field but is
less influenced by density or driving energy. If the magnetic field
goes up by a factor of 4 the time to stall drops by 32!

Nevertheless, the time to stall is long. If we insert
${L_{37}}={3}$, ${n_0}={0.3}$, and ${B_{-6}}={2}$
(note that this is the ambient field, not the compressed field, and
that we use the regular component of the field, ignoring the random
component), then we obtain ${t_{max}}={89}$~Myr, far longer than the
age of OCl~352. If the field is as high as ${B_{-6}}\approx5$, this
time drops to about 9~Myr, a more reasonable value.
The radius of the bubble at stall is
\begin{equation}
{R_{max}}={{2.2 \times 10^{3}}\thinspace{L_{37}^{1/2}}\thinspace{n_{0}^{1/4}}
\thinspace{B_{-6}^{-3/2}}\thinspace{\rm{pc.}}}
\end{equation}
For the situation considered above the stall radius is $\sim$1000~pc
(for ${B_{-6}}={2}$) and $\sim$250~pc (for ${B_{-6}}={5}$).
G134.4+3.85 is nowhere near this size, and has not stalled.

However, bubbles do in time stall.  Contraction follows stall, and
occurs mostly in the equatorial region, so that old bubbles become
quite elongated in the direction of the field (de Avillez \&
Breitschwerdt 2005). However, we have just demonstrated that
G134.4+3.85 has not stalled, so the elongation cannot be the product
of contraction, but must be the product of an ambient field that
already has a strong vertical component.  We therefore propose a model
in which G134.4+3.85 is driven more by pre-existing magnetic field
structure than by pre-existing density structure; this is consistent
with the ideas expressed by de Avillez \& Bretischwerdt (2005).

Ferri{\`{e}}re et al. (1991) describe a substantial thickening of the
wall of a bubble formed in a magnetized medium. In their calculation
for ${L_{37}}={3.2}$, ${n_0}={0.32}$, and ${B_{-6}}={3}$, plausible
values for G134.4+3.85, the equatorial shell thickness is about 20\%
of the radius. From our observations the value of ${\Delta}R/R$ is
between 12 and 24\%, quite compatible with their model.

Pursuing this idea further, we suggest another possible explanation
for the observed gaps in the shell. In the analytical models, the
shell thickness depends on location on the shell. The above estimate
refers to the magnetic equator; at the magnetic poles the shell
becomes very thin. We speculate that this effect might be related to
the apparent break in the top of the shell. It is conceivable that it
is not a break at all, but simply a thin spot.  We note that it is
exactly at such a thin spot at the magnetic pole that one would expect
such a shell to break.

As well as these magnetic field arguments, we find morphological
evidence that supports a scenario with multiple star forming
epochs. Near the cluster there is relatively intense $\sim$22 K
emission at 1420 MHz. Around the cluster, but not symmetrically
centered on it, is the W4 loop, a 1\arcdeg\ ($\sim$40 pc) loop visible
in radio continuum, infrared, and \ha\ emission with shell-like
structure. This loop does not have a constant limb brightness and is
interpreted as having ``broken out'' in places (Terebey et al. \
2003).

Above $b +1.29\arcdeg$, the 1420 MHz intensity drops from 8.8 K to
$\le$6 K. Between $b \approx +1.5\arcdeg$ and $b \approx +3.2\arcdeg$,
the intensity decreases only slightly, from $\sim$6 K to $\sim$5.2K,
after which one finds an arc above which the intensity drops quickly
to $\sim$4.75 K in the low emission region (see
Figure~\ref{fig:Ivsb_Ivsl}. Further above this we find the upper limb
of G134.4+3.85. We postulate that this pattern is the result of a
series of star forming epochs in the region. Detailed theoretical
modeling is beyond the scope of this paper, but is needed to test
whether this postulate is physically reasonable.

\subsection{The halo connection}
\label{sec:halo}

DTS97 state that the O-type stars in OCl 352 have a total output of
$\sim2.3\times 10^{50}$ photons/s. Terebey et al.\ (2003) conclude
that 40\% $\pm$ 16 \% ($\sim9.2 \times 10^{49}$ photons/s) of this
ionizing radiation escapes to heights above 60 pc. They suggest that
these photons are available to ionize the diffuse gas and the upper
Galactic halo. Our observations suggest that some of these photons are
ionizing the gas in the upper portion of the shell.

DTS97 predict that the \ha\ surface brightness of the upper portion of
the shell should be $\sim$20 R, which is well in excess of the 0.7 to
6.7 R that is observed. They propose that two small molecular clouds
located a little above the W4 loop are ``shadowing'' the upper portion
of the shell, accounting for the low observed values of the emission
measure. In contrast, Terebey et al.\ (2003) find that fewer photons
than DTS97 predict are absorbed by the molecular clouds. A schematic
diagram summarizing this picture is provided in
Figure~\ref{fig:ion_schema}.

We adopt the Terebey et al.\ picture that 40\% of the ionizing photons
emitted by OCl 352 are not absorbed by the W4 loop. In light of the
\hi\ and continuum data presented here, which are reminiscent of the
ionization modeling by BJM99, we propose that some of these photons
are ionizing G134.4+3.85 while others escape through the breaks in
G134.4+3.85's walls, and yet others simply pass through the thin,
fully-ionized upper portion of the shell.

The average value of the observed \ha\ surface brightness for the four
points, H through K (see Figure~\ref{fig:contours}) is 4 R. If we
compare this to the predicted 20 R (DTS97) we find that $\sim$20\%
($\sim1.8 \times 10^{49}$ photons/s) of the photons escaping from the
lower loop ionize G134.4+3.85 and the remaining $\sim$80\%
($\sim7.2 \times 10^{49}$ photons/s) escape.

In summary, the cluster produces of total flux of $\sim 2.3 \times
10^{50}$ photons/s, of which 40\% ($\sim9.2 \times 10^{49}$ photons/s)
escape the W4 loop according to Terebey et al.\ (2003). Of the photons
escaping the W4 loop, $\sim$20\% ($\sim1.8\times10^{49}$ photons/s)
are needed to produce the observed ionization in the upper shell of
G134.4+3.85; this is 8\% of the total flux. The amount of this ionizing flux is less than the uncertainty in the estimate of the number of photons escaping from the loop. This helps emphasize that the majority of these loop escaping photons may leave G134.4+3.85 through
the breaks and through the thin, fully-ionized upper shell.

Reynolds et al.\ (2001) estimate that the total hydrogen ionization
rate for the large \ha\ loop observed in the WHAM data is $3 \times
10^{48}$ photons/s, or $\sim$1.3\% of OCl 352's total output. However,
given that the loop subtends $\sim$0.4 sr as seen from the cluster,
approximately 40\% of the cluster's luminosity ($\sim9.2 \times
10^{49}$ photons/s) must escape if such escape is isotropic. This is
reasonably consistent with our estimate, and we conclude that $8 \pm 1
\times 10^{49}$ photons/s) escape into the halo.

\section{Conclusions}
\label{sec:conclu}

Compelling evidence has been presented for the existence of a
fragmented superbubble that is in the process of evolving into a
chimney. Based on low latitude \hi\ data, the structure, named here
G134.4+3.85, was originally postulated to be a fully-formed
chimney. Later, \ha\ data suggested that it was closed near $b =
+7\arcdeg$. The radio continuum data presented in this paper show a
structure similar to that seen in \ha, though two breaks, one at the
top and one in the low longitude wall, are apparent in the combined
data. Polarized intensity images, showing depolarization extending
from the W4 \hii\ region up the walls of the superbubble, confirm that
the observed continuum emission and \ha\ emission are at the same
distance as the W4 region. In contrast, the \hi\ data make it clear
that there is no corresponding neutral shell. We speculate that one of the breaks could be associated with the thinning of the ionized shell at a magnetic pole. However the 2 breaks in the ionized skin of the bubble, and the lack of a closed neutral HI shell, are similar to features in computational simulations (in the literature) of superbubbles which are evolving into chimneys.  In these simulations fragmentation can be caused when magnetohydrodynamic instabilities mix the magnetized shell with the ambient medium.

The fraction of ionizing photons emitted by OCl 352 that are not
absorbed by the W4 loop is more than sufficient to account for the
continuum emission from G134.4+3.85. The excess can escape both
through the two breaks observed in the shell as well as through the
thin, fully ionized upper portion of the shell.

G134.4+3.85 is $\sim$165 pc wide and extends $\sim$240 pc above the
mid-plane of the Galaxy, reaching beyond 135 pc which is the one-$\sigma$
height of the global Gaussian \hi\ distribution. While this is not into the
Galactic halo, the photons from OCl 352 can contribute to the
ionization of the Reynolds layer of the Galaxy and are likely the main
source of ionization of the large \ha\ loop seen in the WHAM data.

\acknowledgements{}
The authors would like to thank R. Kothes for assistance with the data
processing and A. Gray and J.C. Brown for many helpful discussions.

This work was supported by the National Science and Engineering
Research Council of Canada. The Dominion Radio Astrophysical
Observatory Synthesis Telescope is a national facility operated by the
National Research Council of Canada. The Canadian Galactic Plane
Survey is a Canadian project with international partners, and is
supported by a grant from the Natural Sciences and Engineering
Research Council of Canada. Data from the Canadian Galactic Plane
Survey are publicly available through the facilities of the Canadian
Astronomy Data Centre (http://cadc.hia.nrc.ca). The Simbad database is
operated at the Centre de Donn\'ees astronomiques de Strasbourg (CDS),
Strasbourg, France.

\clearpage

\figcaption[]{G134.4+3.85 in \hi\ centered at --43.40 \kms with a channel width of 2.64 \kms. The ``*'' marks the approximate location of OCl 352.
\label{fig:HI-43.40_general}}

\figcaption[]{The G134.4+3.85 in \hi\ centered at --43.40 \kms with a channel width of 2.64 \kms. The grey scale was chosen to highlight the faint upper latitude structures. The ``*'' marks the approximate location of OCl 352.
\label{fig:HI-43.40_faint}}

\figcaption[]{G134.4+3.85 in \hi\ centered at --40.11 \kms with a channel width of 2.64 \kms. This channel represents the upper limit where the structure of G134.4+3.85 may still be easily identified.
\label{fig:HI-40}}

\figcaption[]{G134.4+3.85 in \hi\ centered at --48.35 \kms with a channel width of 2.64 \kms. This channel represents the lower limit where the structure of G134.4+3.85 may still be easily identified.
\label{fig:HI-48}}

\figcaption[]{408 MHz (74 cm) continuum image linearly scaled to show the faint emission. The resolution is $\sim$3.7\arcmin. The white contours are at levels of 120 K and 65 K. The grey contour is at 80 K. The latitude and longitude scales shown have units of degrees. The ``*'' marks the approximate location of OCl 352.
\label{fig:408}}

\figcaption[]{1420 MHz (21 cm) Stokes I continuum image linearly scaled to show the fainter regions and convolved to match the lower resolution of the 408 MHz data ($\sim$3.7\arcmin). White contours are set at levels of 16 K and 6.5 K and grey contours are set at 8.5 K and 5.75 K. The latitude and longitude scales shown have units of degrees. The ``*'' marks the approximate location of OCl 352. 
\label{fig:StokesI}}

\figcaption[]{1420 MHz (21 cm) Stokes Q image (top) and Stokes U image (bottom) convolved to a resolution of 5\arcmin. Rings are artefacts due to the bright sources W3 (center) and 3C~58 (right). The latitude and longitude scales shown have units of degrees. 
\label{fig:QU}}

\figcaption[]{Polarized intensity image (top) and polarization angle map (bottom) at a resolution of 5\arcmin. In the PI image, grey scale is in Kelvin and runs from 0 K (white) to 0.140 K (black) such that highly polarized regions appear black and depolarized regions show white. Note the depolarization along the ``walls'' of the region extending vertically upwards at $\ell \approx 136\arcdeg$ and at $\ell \approx 133.5\arcdeg$ visible as regions of low polarized intensity and as a region having a small ``cell size''.  The ``wishbone'', with approximate center coordinates $\ell = 134.9\arcdeg, b = +7.17\arcdeg$, shows grey in PI and is more prominent in $\psi$ as a region of smoothly varying polarization angle (see \S\ref{sec:wishbone}). Note also, the bright polarized  ``knot'' centered at approximately $\ell = 134.5\arcdeg, b = +2.58\arcdeg$ and showing as black in PI (see \S\ref{sec:comma}). The interesting lenticularly shaped feature in $\psi$ centered at approximately $\ell = 137.5\arcdeg, b = +1\arcdeg$ was discussed as likely being due to a foreground object by Gray et al. (1998, 1999). The latitude and longitude scales shown have units of degrees. 
\label{fig:PIPang}}

\figcaption[]{The following  datasets have been assigned the colours listed in parentheses: 1420 MHz (turquoise), 408 MHz (purple), \ha\ (red), and 60 $\mu$m IRAS (yellow). Before combining, the intensities in each set were  logarithmically scaled in order to display faint emission. Once colourized, the images were combined. The technique used is analogous to overlaying transparent film images and thus allows details from each dataset to be displayed in the final colour image. These features are shown here for positional reference. The supernova remnant, HB3 shows a distinct purplish colour indicating dominant emission at 1420 and 408 MHz. Regions showing prominently as yellow have high dust emission. 
\label{fig:colour}}

\figcaption[]{
 Greyscale version of the image as described in Figure~\ref{fig:colour} with labels. The labels identify objects with catalogue designations as well as prominent features visible in the data sets. G134.4+3.85 traces the continuous structure apparent in these data. The ``wishbone'' is a feature visible in the polarization data and the V-shaped and other filaments are \hi\ spectral line features (NTD96). The ``*'' marks the approximate location of OCl 352.
\label{fig:labelled}}

\figcaption[figures/coldens_26m_ch60to62.ps]{\hi\ column density near G134.4+3.85 for the velocity interval --41.76 \kms to --45.05 \kms. The column density was evaluated from the low spatial resolution, 26-m Telescope data which cover a greater latitude range than the Synthesis Telescope data. The latitude of the OCl 352 cluster is indicated by the arrow. The dotted line is an exponential fit to the decrease in $N_h$. The implied scale height is approximately 140 pc. A distance of 2.35 kpc is assumed. 
\label{fig:Nvsz}}

\figcaption[]{G134.4+3.85 and lines tracing the contours of higher emission showing the fork and the two breaks. The datasets are described in Fig.~\ref{fig:colour}. The points labelled alphabetically are regions measured in \ha\ emission by DTS97. These \ha\ features are discussed in \S\ref{sec:chimney}. The ``*'' marks the approximate location of OCl 352.
\label{fig:contours}}

\figcaption[]{a) Plot of intensity vs Galactic latitude in the 1420 MHz data at a fixed longitude of 134.68\arcdeg. b) Plot of intensity vs Galactic longitude in the 1420 MHz data at a fixed latitude of +4.011\arcdeg. 
\label{fig:Ivsb_Ivsl}}

\figcaption[]{Schematic diagram summarizing the results of DTS97, Terebey et al.\ (2003) and Reynolds et al.\ (2001) regarding the number of ionizing photons that can escape into the Galactic halo. DTS97 predict that the upper shell should have an intensity of 20 R but it was observed to have $\sim$1/5 this intensity. The region dubbed as G134.4+3.85 includes the filaments above a b=+1\arcdeg and extending up to b=+6\arcdeg. This is defined by the egg shaped structure outlined in Figure~\ref{fig:labelled}. The dark and light grey lines schematically represent the varying intensity of the emission.
\label{fig:ion_schema}}

% \newpage
% \plotone{f1.jpg}
% \newpage
% \plotone{f2.jpg}
% \newpage
% \plotone{f3.jpg}
% \newpage
% \plotone{f4.jpg}
% \newpage
% \plotone{f5.jpg}
% \newpage
% \plotone{f6.jpg}
% \newpage
% \plotone{f7.jpg}
% \newpage
% \plotone{f8.jpg}
% \newpage
% \epsscale{1.0}
% \plotone{f9.jpg}
% \newpage
% \plotone{f10.jpg}
% \newpage
% \plotone{f11.jpg}
% \newpage
% \plotone{f12.jpg}
% \newpage
% \plotone{f13.jpg}
% \newpage
% \epsscale{1.0}
% \plotone{f14.jpg}

\clearpage

\begin{deluxetable}{c c c c c c}
\tablewidth{0pt}
\tablecaption{Center coordinates and approximate widths of ``breaks'' in G134.4+3.85.
\label{tbl:breaks}}
\tablehead{
\colhead{} & \colhead{1420 MHz} & \colhead{408 MHz} & \colhead{\ha} & \colhead{width$^a$}  \\
}
\startdata
upper break & 134.77\arcdeg, +5.58\arcdeg & 134.12\arcdeg,
+5.82\arcdeg & 135.08\arcdeg, +5.97\arcdeg & 0.5\arcdeg (20 pc) \\ 
side break & 133.47\arcdeg, +4.36\arcdeg & 132.84\arcdeg, +3.73\arcdeg
& 133.56\arcdeg, +4.15\arcdeg & 1.2\arcdeg (49 pc) \\ 
\enddata
\tablenotetext{[a]}{The values in parentheses are for an assumed distance of 2.35 kpc.}
\end{deluxetable}

\end{document}